\documentclass{article} 
\usepackage{iclr2027_conference,times}

\usepackage{amsmath,amsfonts,bm}

\def\eqref#1{equation~\ref{#1}}

\def\1{\bm{1}}

\DeclareMathAlphabet{\mathsfit}{\encodingdefault}{\sfdefault}{m}{sl}
\SetMathAlphabet{\mathsfit}{bold}{\encodingdefault}{\sfdefault}{bx}{n}

\usepackage{hyperref}
\usepackage{url}
\usepackage{graphicx}
\usepackage{subcaption}
\usepackage{tabularx}
\usepackage{booktabs} 
\title{VisFlow: Interpretable Flow Matching for Visibility Reconstruction, Characterization, and Recognition}

\author{Rahul Moorthy and Volkan Isler
}

\iclrfinalcopy 
\begin{document}

\maketitle
\lhead{Preprint}

\newcommand{\vtxt}[1]{\textcolor{red}{{\em #1}}}
\newcommand{\rtxt}[1]{\textcolor{blue}{{\em #1}}}

\newcommand{\ourprob}{\textit{Visibility Reconstruction}}
\newcommand{\ourproball}{\textit{Visibility Characterization}}
\newcommand{\ourprobrec}{\textit{Visibility Recognition}}

\newcommand{\ourmethod}{\textbf{Cont-VisDiff}}
\newcommand{\ourmethodsmall}{Cont-VisDiff}

\newcommand{\ourmethodnew}{\textbf{VisFlow}}
\newcommand{\ourmethodnewsmall}{VisFlow}

\begin{abstract}

Methods for learning continuous representations, such as latent manifolds, are fundamental and ubiquitous in modern machine learning. Yet many geometric objects combine continuous shape information with discrete combinatorial structure. This combination presents a fundamental challenge for representation learning because small geometric perturbations can induce abrupt changes in the underlying discrete representation. As a result, the smoothness assumptions that make continuous representations effective no longer hold. Polygons provide a canonical example. Their vertex locations vary continuously, but infinitesimal changes in geometry can dramatically alter the associated visibility graph. Existing methods address the inverse mapping from visibility graphs to polygon geometry using vertex-, triangulation-, or signed-distance-function (SDF)-based representations. Vertex- and triangulation-based approaches construct polygons through explicit geometric operations. However, they are less effective at modeling the relationship between geometry and visibility. SDF-based approaches capture this relationship more naturally. Their intermediate states, however, do not correspond to explicit stages of polygon construction.

\par We present \ourmethodnew{}, a novel, interpretable flow formulation that combines continuous SDF representations with explicit polygon evolution. Our method models polygon generation as rectified flow in the latent space of continuous SDFs. Standard rectified flow typically constructs linear paths between samples from an unstructured source distribution and the target distribution. The resulting intermediate states are not tied to stages of polygon formation. In contrast, \ourmethodnew{} learns approximately linear latent paths to represent explicit stages of polygon growth along SDF normals. We evaluate \ourmethodnew{} on \ourprob{} and \ourproball{}. \ourprob{} generates one polygon consistent with a visibility graph \(G\), while \ourproball{} characterizes all such polygons. \ourmethodnew{} maintains \textbf{20\%} more visibility relations on \ourprob{} and improves coverage by \textbf{30\%} on \ourproball{} compared to state-of-the-art methods. We further show that \ourmethodnew{} generalizes to \ourprobrec{}, where the input graph is not guaranteed to be valid. We also demonstrate that \ourmethodnew{} extends to a real-world protein reconstruction problem by applying its trajectory-supervision formulation to generate protein structures from contact graphs.
\end{abstract}
\section{Introduction}
Learning the mapping from a graph to its geometric realization arises in problems such as molecular conformation generation, spatial layout reconstruction, and mesh generation. Existing approaches often use continuous geometric attributes, such as coordinates or distances, alongside graph connectivity to represent the underlying geometry. Local perturbations produce small changes in these attributes, keeping nearby geometric realizations close in the representation space. The resulting continuity forms a smooth geometric manifold that learning models can exploit. Purely discrete graph representations lack an analogous continuous structure. Small geometric perturbations can change several edge relations simultaneously, causing similar realizations to map to substantially different graph representations. Therefore, learning the inverse mapping from a discrete graph to its geometric realization is more difficult because nearby geometries are not necessarily nearby in the representation space.

In this paper, we study one such discrete representation: the visibility graph of a polygon. Two polygon vertices are connected by an edge if the line segment joining them lies entirely within the polygon. Small perturbations of the polygon can change several visibility relations simultaneously. Prior work has explored triangulation, vertex-based, and signed distance function (SDF) representations for learning polygon realizations from visibility graphs. Triangulation and vertex-based representations explicitly model geometric construction. These representations have shown promising results in mesh generation, where continuous geometric information is available. However, they struggle to preserve the target visibility structure when conditioned only on a discrete visibility graph. A visibility graph contains no distance information between vertices. In contrast, SDF representations provide a continuous distance representation of the polygon and capture the relationship between polygon geometry and visibility more effectively. Existing SDF-based methods start from Gaussian noise and perform diffusion directly in grid space. Their generative process therefore does not explicitly model polygon formation. Our work asks whether the construction-based formulation of geometric representations can be incorporated into SDF-based generative modeling while retaining the strong relationship between SDFs and visibility?

\begin{figure}[h!]
    \centering
    \includegraphics[
    width=\textwidth,
    trim=1cm 5cm 0.5cm 5.5cm,
    clip
]{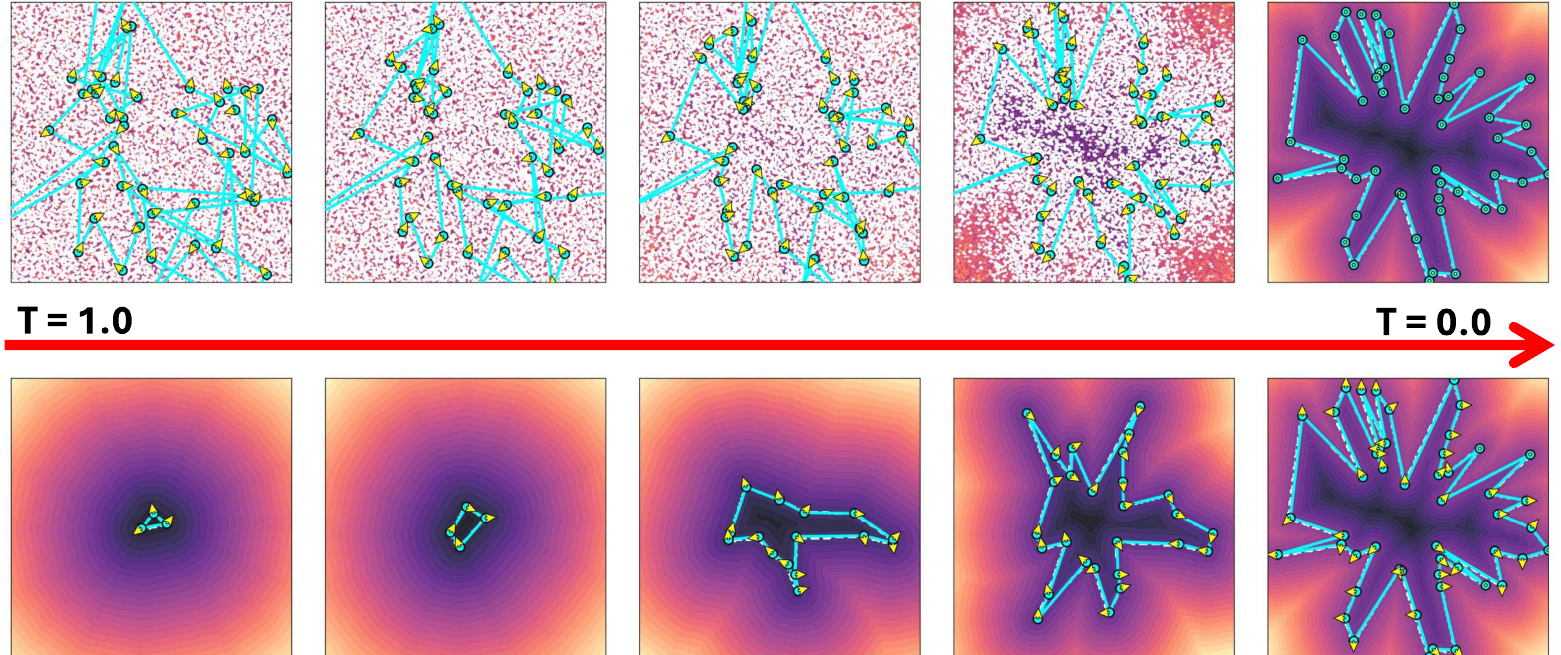}
   \caption{\textbf{Rectified Flow vs polygon evolution.}
    \textbf{Top:} SDF rectified flow starts from Gaussian noise and produces intermediate grid states that are not explicitly tied to stages of polygon formation.
    \textbf{Bottom:} Our formulation starts from a triangle and evolves through intermediate polygon states toward the final polygon.}
    \label{fig:intro_figure}
\end{figure}

We present \ourmethodnew{}, a geometrically meaningful flow-matching model that combines a continuous SDF representation with explicit polygon evolution. Standard flow matching typically learns trajectories from an unstructured source distribution, such as Gaussian noise, to the target data distribution. The intermediate states along such trajectories are not tied to stages of polygon formation. Figure~\ref{fig:intro_figure} illustrates the contrast between such unstructured intermediate states and the explicit polygon evolution used in \ourmethodnew{}. \ourmethodnew{} instead learns an SDF latent representation from polygon trajectories that evolve from an initial triangle to the target polygon. The latent representation is trained so that the intermediate polygon states lie along approximately linear paths. We then use rectified flow to learn the transformation from the initial triangle latent to the target polygon latent, conditioned on the visibility graph. We additionally learn a differentiable Visibility Prediction Network that provides visibility supervision during training and guides optimization of the generated polygon toward the target visibility graph during inference. We further apply the same trajectory-supervision formulation to protein structure generation from contact graphs, demonstrating its use with a different underlying structural evolution process.

Our contributions are as follows:

\begin{itemize}
        \item We present \ourmethodnew{}, a geometrically meaningful flow-matching model that combines a continuous SDF representation with explicit polygon evolution for visibility-conditioned polygon generation. We also introduce a Visibility Prediction Network that provides differentiable visibility supervision during training and enables visibility-guided optimization during inference. \ourmethodnew{} improves F1-score by \textbf{20\%} on \ourprob{} and improves coverage by \textbf{30\%} on \ourproball{} compared with state-of-the-art methods.

\item We extend the trajectory-supervision formulation of \ourmethodnew{} to protein structure generation from contact graphs.

\item We present a new dataset for \ourprob{} with \textbf{22\%} greater shape diversity than the state-of-the-art dataset.

    \end{itemize}

\section{Related Work}

We categorize related work into three directions: (i) visibility graph reconstruction, recognition, and characterization, (ii) interpretable flow matching, and (iii) polygon propagation methods.

\textbf{Visibility Graph Reconstruction, Recognition, and Characterization:} Prior work has developed reconstruction and recognition algorithms for restricted polygon classes, including staircase, tower, anchor, spiral, terrain, and convex fan polygons~\citep{abello1995visibility,colley1997visibility,boomari2016visibility,everett1990recognizing,silva2020visibility}. However, the general visibility-graph recognition problem remains open~\citep{ghosh2013unsolved}. More recently, learning-based methods have modeled the relationship between polygon geometry and visibility. VisDiff~\citep{mahesh2026visdiff} showed that diffusion through an intermediate SDF representation better preserves visibility graphs than GNN- and mesh-based alternatives~\citep{cui2023mgnngraphneuralnetworks,chen2024meshanythingartistcreatedmeshgeneration,lionar2025treemeshgpt}. However, its coarse grid-based SDF restricts representable polygon geometry, and its intermediate diffusion states do not correspond to a geometry-driven process of polygon formation. We address these limitations using a continuous SDF representation and explicit polygon-evolution supervision.

\textbf{Flow Matching:} Flow matching learns a continuous-time vector field that transports samples from a source distribution to the data distribution~\citep{lipman2022flow}. Rectified Flow~\citep{liu2022flow}, a flow-matching formulation, encourages approximately straight source-to-target trajectories. Standard formulations typically use unstructured source distributions such as Gaussian noise, so their intermediate states are not tied to the evolution of the generated structure. Recent work has explored giving flow trajectories physical meaning. Pivi et al.~\citep{pivi2025flow} use transitions between equilibrium states of an Ising system, while Singh et al.~\citep{pmlr-v300-singh26a} combine flow matching with partially known ODE/PDE dynamics. These approaches incorporate the physical process by changing the dynamics that the flow is trained to follow. In contrast, we use a meaningful geometric evolution to supervise the latent space without replacing the efficient source-to-target transport used for generation by flow matching.

\textbf{Polygon Propagation Approaches:} Polygon propagation defines controlled geometric shape evolution. Grassfire propagation moves the polygon boundary inward along SDF normals~\citep{blum1967transformation}, while Bruckstein et al.~\citep{bruckstein1995evolutions} update vertices based on their neighbors. Learning-based methods have also used polygon evolution for shape fitting: Deep Snake~\citep{peng2020deep} iteratively deforms a polygon toward an object boundary, while Deep Extreme Level Set Evolution~\citep{wang2019object} learns implicit-contour evolution for segmentation. In these methods, the target shape is given and evolution is used to fit its boundary. In contrast, we start from a visibility graph and use polygon evolution to supervise generation of a polygon consistent with the input visibility graph.

\section{Problem Formulation}

We focus on simple (no boundary crossing), simply connected polygons (i.e., without holes). Let $X \in \mathbb{R}^{N \times 2}$ denote the $N$ vertex coordinates of a polygon $P$, and let $G \in \mathbb{R}^{N \times N}$ denote its visibility graph represented as an adjacency matrix. The function $V(P)$ computes the visibility graph of $P$. A visibility graph $G$ is valid if there exists at least one polygon $P$ such that $V(P)=G$.

\textbf{\ourprob{}}: Given a valid visibility graph $G$, generate a polygon $P$ such that $V(P) = G$.

\textbf{\ourproball{}}: given a valid visibility graph $G$, generate all simple polygons $P$ such that $V(P) = G$.

\textbf{\ourprobrec{}}: Given a graph $G$, determine whether there exists a polygon $P$ such that $V(P) = G$.

\begin{figure}[h!]
    \centering
    \includegraphics[
    width=\textwidth,
    trim=0 9.5cm 0 1.5cm,
    clip
]{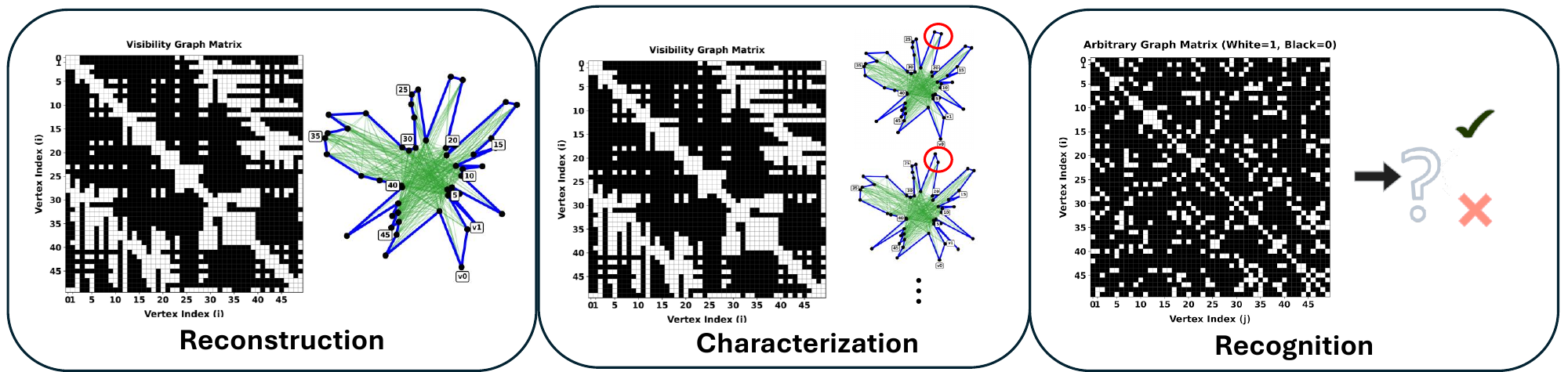}
    \caption{\textbf{Visibility graph problems.} Reconstruction finds a polygon consistent with a given visibility graph, characterization finds all polygons corresponding to the same visibility graph (The red circle shows the perturbations but still the same visibility graph), and recognition determines whether an arbitrary graph is a valid visibility graph.}
    \label{fig:visibility_problems}
\end{figure}
\vspace{-1.5em}
\section{Method}

\begin{figure}[h!]
    \centering
    \includegraphics[width=\columnwidth]{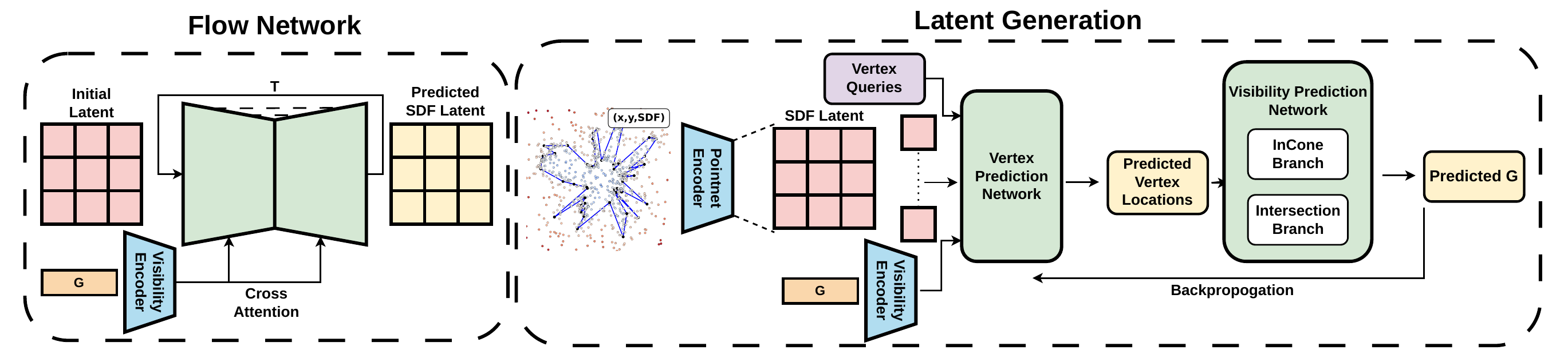}
\caption{\textbf{Overview of \ourmethodnewsmall{}.}
\textbf{Left:} The Flow Network evolves an initial triangle latent into a target polygon latent using a time-conditioned U-Net conditioned on the target visibility graph.
\textbf{Right:} The SDF Latent Generator encodes a continuous SDF into spatially localized latent features and decodes polygon vertices conditioned on the visibility graph. The Visibility Prediction Network provides differentiable visibility supervision.
\textbf{Inference:} Given a target visibility graph $G$, we sample and encode an initial triangle, evolve its latent using the learned flow, decode the final latent into a polygon, and refine its vertices using the Visibility Prediction Network to better match $G$.}
    \label{fig:architecture}
\end{figure}

We now detail the architecture of \ourmethodnewsmall{}. The central idea is to encode the signed distance function (SDF) of a polygon into spatially localized latent features and learn a latent representation in which stages of polygon formation follow approximately linear trajectories. We construct these trajectories by applying grassfire evolution~\citep{blum1967transformation} to progressively collapse each polygon to a triangle and then reversing the resulting trajectory. During sampling, we draw an initial triangle and evolve its latent representation to generate a polygon conditioned on the target visibility graph. \ourmethodnewsmall{} consists of three main components: an \textbf{SDF Latent Generator} that learns spatially localized SDF features and the corresponding polygon-formation trajectories, a \textbf{Visibility Prediction Network} that provides differentiable visibility supervision, and a \textbf{Flow Network} that models polygon generation conditioned on the target visibility graph. We train the components sequentially: first the Visibility Prediction Network, then the SDF Latent Generator with the visibility network frozen, and finally the Flow Network with both networks frozen. Figure~\ref{fig:architecture} provides an overview of the complete architecture. Grassfire evolution~\citep{blum1967transformation} and the visibility formulation motivating our Visibility Prediction Network are detailed in Appendix~\ref{app:prelim}.

\subsection{SDF Latent Generator}

The SDF Latent Generator learns a compact representation of the continuous SDF while preserving the progression of polygon formation. We use a continuous SDF to avoid dependence on a fixed grid resolution. Direct generative modeling of the continuous SDF would require the flow model to operate over dense spatial SDF samples at every integration step, resulting in high computational and memory costs. We therefore encode the SDF into a compact latent representation. We define polygon-formation trajectories using grassfire flow~\citep{blum1967transformation}. The inward evolution is stopped when only a triangle remains, the simplest non-degenerate polygonal state. We then reverse the trajectory so that polygon generation proceeds from the triangle to the complete polygon. We normalize trajectory time to \(t\in[0,1]\), with \(t=0\) at the triangle and \(t=1\) at the complete polygon. The resulting grassfire trajectories depend on polygon geometry and therefore follow different deformation paths across shapes. A common latent parameterization is needed to represent these trajectories consistently across polygons. We therefore learn a latent representation in which the trajectories follow approximately straight paths between their endpoint latents. Intermediate grassfire states supervise points along each latent path so that linearly interpolated latents decode to the corresponding stages of polygon formation when conditioned on their visibility graphs.

The straight-line latent trajectories are learned using an encoder--decoder architecture. The encoder represents the input SDF using \((x,y,\mathrm{SDF})\) point samples. We adopt the PatchNets formulation~\citep{tretschk2020patchnets} and represent the encoded SDF using spatially localized patch latents over \([-1,1]^2\). PointNet~\citep{qi2017pointnet} maps each SDF sample to a feature representation, which is aggregated within its corresponding spatial patch to produce the patch latent. A positional embedding of each patch center encodes its spatial location, and a Transformer encoder exchanges information across patches to produce a context-aware SDF representation. The decoder uses learnable vertex queries to recover polygon vertex locations. The visibility graph \(G\) is processed to produce one visibility embedding per vertex, which is concatenated with the corresponding vertex query. A Transformer decoder then cross-attends the resulting vertex representations to the context-aware patch latents and predicts the polygon vertex locations. Additional architecture details are provided in Appendix~\ref{app:sdf_architecture}.

We train the SDF Latent Generator using reconstruction, alignment, and regularization losses to preserve polygon geometry and visibility while encouraging linear latent trajectories. The overall objective is
\[
\mathcal{L}_{\mathrm{latent}}
=
\mathcal{L}_{\mathrm{rec}}
+
\lambda_{\mathrm{ali}} \mathcal{L}_{\mathrm{ali}}
+
\lambda_{\mathrm{reg}} \mathcal{L}_{\mathrm{reg}}.
\]
The alignment loss is
\[
\mathcal{L}_{\mathrm{ali}}
=
\left\|
Z_t^{\mathrm{enc}}
-
\operatorname{sg}\!\left(
\tilde{Z}_t
\right)
\right\|_2^2,
\qquad
\tilde{Z}_t
=
(1-t)Z_{\mathrm{start}}
+
tZ_{\mathrm{end}}.
\]
The reconstruction loss \(\mathcal{L}_{\mathrm{rec}}=\mathcal{L}_{\mathrm{coord}}+\mathcal{L}_{\mathrm{vis}}\) combines geometric and visibility supervision. The coordinate loss \(\mathcal{L}_{\mathrm{coord}}\) measures the mean squared error between predicted and ground-truth vertices at the start, end, and intermediate states decoded from \(\tilde{Z}_t\). Each intermediate state is conditioned on the visibility graph of the corresponding ground-truth grassfire state so that the linearly interpolated latents decode to the corresponding stages of polygon formation. Coordinate reconstruction alone does not enforce the desired visibility relations, so we additionally use \(\mathcal{L}_{\mathrm{vis}}\), defined as the binary cross-entropy between the ground-truth visibility graph and the prediction from the frozen Visibility Prediction Network. We next use the alignment loss \(\mathcal{L}_{\mathrm{ali}}\) to encourage the encoded grassfire states to follow the linear path between the endpoint latents. The loss encourages the encoded grassfire state \(Z_t^{\mathrm{enc}}\) to match the corresponding point \(\tilde{Z}_t\) on this path, while \(\operatorname{sg}(\cdot)\) prevents the endpoint latents from being updated by the alignment objective. Since the alignment loss does not control the magnitude of the latent representation, \(\mathcal{L}_{\mathrm{reg}}\) penalizes the mean squared magnitude of the start, end, and encoded intermediate latents. We set \(\lambda_{\mathrm{ali}}=1.0\) and \(\lambda_{\mathrm{reg}}=10^{-3}\) in all experiments.

\subsection{Visibility Prediction Network}

The Visibility Prediction Network provides a differentiable approximation of polygon visibility. Exact visibility is non-differentiable with respect to polygon vertices, preventing direct visibility supervision during training and gradient-based refinement during inference. We therefore learn a differentiable predictor based on the visibility predicates defined in the preliminaries~(Appendix~\ref{app:prelim}).
\par The network first maps each polygon vertex to a positional embedding. For each candidate segment between vertices \(v_i\) and \(v_j\), the endpoint embeddings are concatenated to form a candidate-segment representation. The intersection branch compares this representation with each non-incident polygon edge and uses an MLP to predict pairwise intersection probabilities. A max operation aggregates these probabilities to obtain \(P_{\mathrm{Intersect}}\), since an intersection with any non-incident edge invalidates visibility. The in-cone branch combines the candidate-segment representation with the neighboring edge representations at each endpoint and uses an MLP to predict the corresponding in-cone probabilities. The two endpoint probabilities are multiplied to obtain \(P_{\mathrm{InCone}}\), since the in-cone condition must hold at both endpoints. The final visibility probability is computed as \(P_{\mathrm{Vis}}=P_{\mathrm{InCone}}(1-P_{\mathrm{Intersect}})\). Both branches are trained using binary cross-entropy with their corresponding ground-truth predicate labels.
\par The pretrained Visibility Prediction Network is frozen during SDF Latent Generator training and used to compute \(\mathcal{L}_{\mathrm{vis}}\), allowing visibility gradients to propagate to the predicted polygon vertices. The same network is used during inference for gradient-based refinement toward the target visibility graph. Table~\ref{tab:loss_ablation} reports the effect of the differentiable visibility loss, Table~\ref{tab:verifier_performance} compares the predicate-based architecture with a standard Transformer-based visibility predictor, and Figure~\ref{fig:verifer_architecture} shows the complete architecture.

\subsection{Flow Network}
The Flow Network models the transition from an initial triangle latent to a target polygon latent conditioned on the target visibility graph. We obtain the endpoint latents \(Z_{\mathrm{start}}\) and \(Z_{\mathrm{end}}\) by encoding the SDFs of the corresponding trajectory endpoints with the pretrained SDF Latent Generator, which remains frozen during Flow Network training. The approximately linear latent trajectories allow us to model these transitions using rectified flow~\citep{liu2022flow}. At time \(t\), we construct the linearly interpolated latent as
\[
\tilde{Z}_t=(1-t)Z_{\mathrm{start}}+tZ_{\mathrm{end}},
\]
with target velocity \(Z_{\mathrm{end}}-Z_{\mathrm{start}}\). We sample \(t\sim\mathcal{U}(0,1)\) and train the Flow Network with the rectified-flow objective
\[
\mathcal{L}_{\mathrm{flow}}
=
\mathbb{E}_{t}\!\left[
\left\|
v_{\theta}(\tilde{Z}_t,t,G)
-
\left(Z_{\mathrm{end}}-Z_{\mathrm{start}}\right)
\right\|_2^2
\right].
\]
The Flow Network predicts the velocity from \(\tilde{Z}_t\), time \(t\), and the target visibility graph \(G\). We parameterize the velocity field using a time-conditioned U-Net following Latent Diffusion~\citep{rombach2022high}. A separate visibility encoder produces one visibility embedding per vertex, which conditions the U-Net through cross-attention. Additional architecture details are provided in Appendix~\ref{app:flow_architecture}.
\paragraph{Sampling.}
Given a target visibility graph \(G\), we sample initial triangles from the valid coordinate range and construct their continuous SDFs. The SDF Latent Generator encodes each triangle SDF to obtain an initial latent. We then integrate the learned velocity field using Euler integration while conditioning on \(G\). The resulting target latent is decoded by the SDF Latent Generator to obtain a polygon. The triangle initialization and approximately linear latent trajectories encourage the intermediate states to follow the corresponding stages of polygon formation. The decoded polygon may not exactly satisfy \(G\), so we refine its vertex locations using binary cross-entropy between the predicted visibility graph and \(G\) through the frozen Visibility Prediction Network. We additionally apply binary cross-entropy to predicted intersections between non-incident polygon edges, using zero as the target to penalize self-intersections. The visibility and intersection losses are summed and propagated through the frozen Visibility Prediction Network to optimize the polygon vertices.

\section{Results}
In this section, we evaluate \ourmethodnewsmall{} on \ourprob{}, \ourproball{}, \ourprobrec{}, and contact-graph protein structure reconstruction. We first introduce the baselines and evaluation metrics, followed by the results for each problem.

\par \textbf{Baselines:} We compare \ourmethodnewsmall{} with Mesh-Generation~\citep{lionar2025treemeshgpt}, C-VAE~\citep{harvey2022conditionalimagegenerationconditioning}, GNN~\citep{cui2023mgnngraphneuralnetworks}, and VisDiff~\citep{mahesh2026visdiff} on \ourprob{}. VisDiff is also used as the baseline for \ourproball{} and \ourprobrec{}. All methods are trained on the training split of our dataset and evaluated on its test split. Details of the dataset generation and splits are provided in the appendix \ref{subsec:app_dataset_generation}

\par
\begin{figure}[h!]
    \centering

    \begin{subfigure}{0.48\textwidth}
        \centering
        \includegraphics[
            height=4cm,
            trim=0 0cm 0 0cm,
            clip
        ]{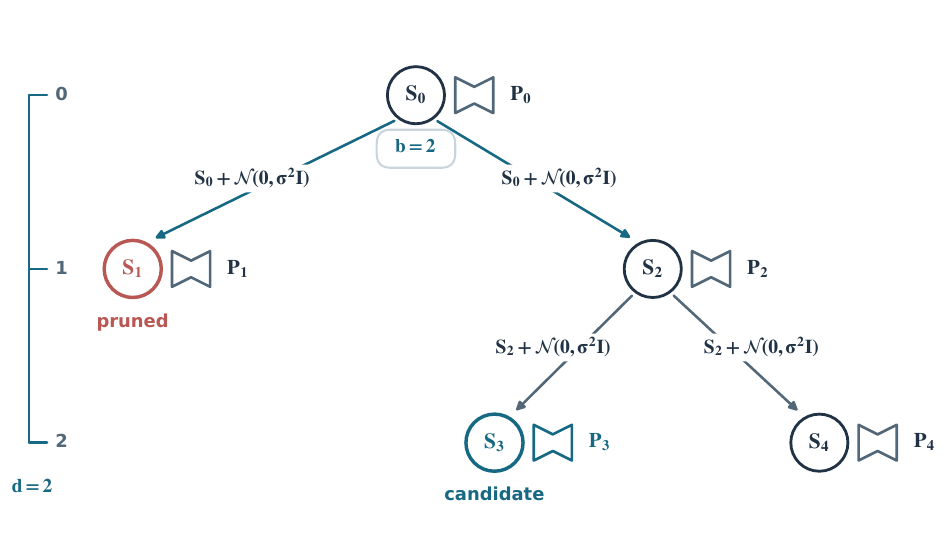}
        \caption{Exploration Tree Generation}
        \label{fig:search_tree}
    \end{subfigure}
    \hfill
    \begin{subfigure}{0.48\textwidth}
        \centering
        \includegraphics[
            height=4cm,
            trim=0cm 1cm 0cm 0cm,
            clip
        ]{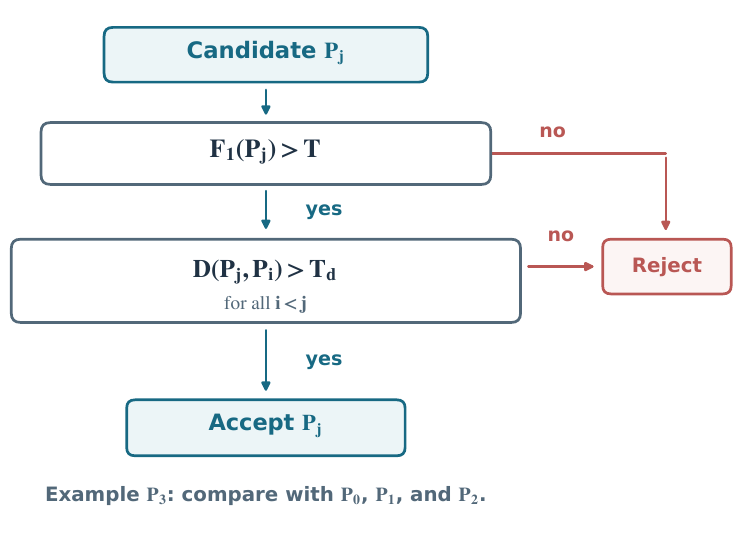}
        \caption{Candidate Acceptance Criteria}
        \label{fig:candidate_acceptance}
    \end{subfigure}

   \caption{\textbf{Coverage:} (a) A learned visibility-graph reconstruction network generates polygon \(P_i\) from source state \(S_i\). Child states are formed as \(S_{\mathrm{child}}=S_{\mathrm{parent}}+\epsilon\), where \(\epsilon\sim\mathcal{N}(0,\sigma^2 I)\). (b) A candidate polygon \(P_j\) is accepted if \(F_1(P_j)>T\) and \(D(P_j,P_i)>T_d\) for every \(i<j\). Here, \(F_1(P_j)\) is the F1 score between the visibility graph of \(P_j\) and the target visibility graph, and \(D(P_j,P_i)\) is the Chamfer distance between the two polygons. Coverage is the ratio of expanded nodes to total nodes, \(N_{\mathrm{expanded}}/N_{\mathrm{total}}\).}
    \label{fig:coverage_figure}
\end{figure}

\textbf{Metrics:} We evaluate \ourmethodnewsmall{} using the metrics defined in VisDiff~\citep{mahesh2026visdiff}. The \ourprob{} problem is evaluated by comparing the visibility graphs of generated and ground-truth polygons as a binary classification task over vertex pairs. We report accuracy, precision, recall, and F1-score, with F1-score used as the primary metric because of class imbalance. The \ourproball{} problem requires measuring how broadly the space of polygons satisfying a target visibility graph \(G\) is covered. Since this space cannot be enumerated exhaustively, we use neighbourhood exploration as a proxy for coverage. We draw a root sample \(S_0\) from the source distribution and generate its corresponding polygon \(P_0\) conditioned on \(G\). We then explore the neighbourhood of \(S_0\) using breadth-first search up to depth \(d\) with branching factor \(b\). Each child sample is obtained by applying scheduled perturbations to its parent and is used to generate a polygon conditioned on \(G\). A node is retained only if its visibility F1-score with respect to \(G\) exceeds \(T\) and its polygon is sufficiently distinct from all previously accepted polygons under the distance threshold \(T_d\). We measure distinctness by aligning the polygons for rotation and translation using Procrustes alignment~\citep{schonemann1966generalized}, uniformly sampling their boundaries, and computing the Chamfer distance between the sampled points. Coverage is defined as the fraction of explored nodes that are retained. We assess diversity by varying \(T_d\). Figure~\ref{fig:coverage_figure} illustrates the coverage metric.
\vspace{-1.5em}
\vspace{-1.5em}
\subsection{Visibility Reconstruction}
\begin{figure*}[h!]
    \centering

    \begin{subfigure}{0.16\textwidth}
        \centering
        \includegraphics[width=\linewidth, trim=0 5cm 22.2cm 5cm,
            clip]{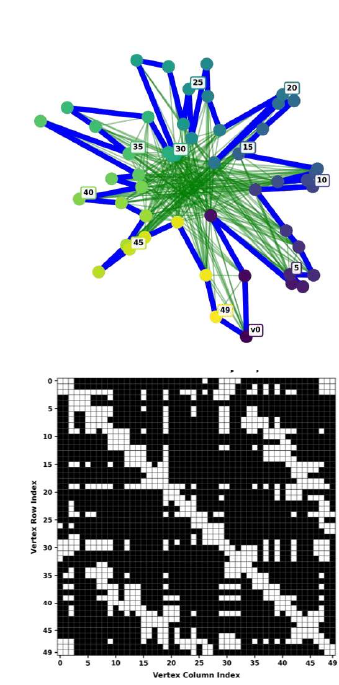}
        \caption{GT}
        \label{fig:subfig1}
    \end{subfigure}
    \hfill
    \begin{subfigure}{0.16\textwidth}
        \centering
        \includegraphics[height=4cm, width=\linewidth, trim=6.5cm 4.7cm 15cm 6cm,
            clip]{Figures/Reconstruction_Qual_1.pdf}
        \caption{\textbf{0.96}}
        \label{fig:subfig2}
    \end{subfigure}
    \hfill
    \begin{subfigure}{0.16\textwidth}
        \centering
        \includegraphics[height = 4.4cm,width=\linewidth, trim=13cm 4.8cm 9cm 5cm,
            clip]{Figures/Reconstruction_Qual_1.pdf}
        \caption{0.75}
        \label{fig:subfig3}
    \end{subfigure}
    \hfill
    \begin{subfigure}{0.16\textwidth}
        \centering
        \includegraphics[width=\linewidth, trim=19.5cm 4.7cm 2.5cm 5cm, clip]{Figures/Reconstruction_Qual_1.pdf}
        \caption{0.66}
        \label{fig:subfig4}
    \end{subfigure}
    \hfill
    \begin{subfigure}{0.16\textwidth}
        \centering
        \includegraphics[height = 4cm, width=\linewidth, trim=6.8cm 5cm 14.5cm 6cm, clip]{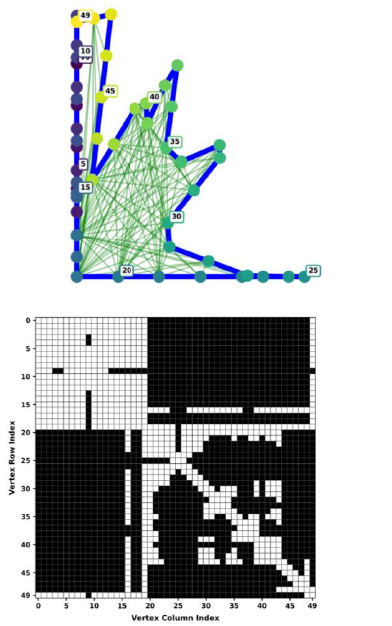}
        \caption{0.63}
        \label{fig:subfig5}
    \end{subfigure}
    \hfill
    \begin{subfigure}{0.16\textwidth}
        \centering
        \includegraphics[height = 4cm,width=\linewidth,trim=13.5cm 5cm 7.5cm 6cm, clip]{Figures/Reconstruction_Qual_2.pdf}
        \caption{0.70}
        \label{fig:subfig6}
    \end{subfigure}

    \caption{\ourprob{} Qualitative Results: (a) GT, (b)\ourmethodnewsmall{}~+~Optimization, (c) VisDiff, (d) GNN, (e) Mesh, (f) C-VAE. The labels below show the F-1 Score comparison with the GT Visibility Graph.}
    \label{fig:recon_qual}
\end{figure*}

We evaluate \ourmethodnewsmall{} on \ourprob{} against the baseline methods. We generate \(N=50\) polygon samples per generative approach and select the polygon whose predicted visibility graph achieves the highest F1-score with respect to the target graph, while the GNN baseline produces a single polygon. Table~\ref{tab:testset_res} shows that \ourmethodnewsmall{} outperforms all baselines. Table~\ref{tab:testset_visdiff_res} further reports its performance on the original VisDiff dataset, and Figure~\ref{fig:recon_qual} provides a qualitative comparison. VisDiff exhibits a substantial performance decline on our 50-vertex dataset compared with its original 25-vertex dataset because its coarse SDF resolution limits faithful representation of polygon geometry at higher vertex counts, whereas the continuous SDF representation in \ourmethodnewsmall{} preserves the geometry. We next examine the source of the performance gains. Table~\ref{tab:approach_ablation_study} shows that inference-time optimization provides additional improvement, while the flow model already outperforms VisDiff and the other baselines without optimization. We then compare \ourmethodnewsmall{} with continuous SDF latent diffusion (\ourmethodsmall{}) in Table~\ref{tab:cont_visdiff_vs_visflow} to separate the effects of polygon propagation and the continuous SDF latent representation. Polygon propagation provides only a small additional improvement on \ourprob{}, indicating that the primary performance gain comes from the continuous SDF representation.
\begin{table}[h!]
    \centering
    \begin{tabularx}{\textwidth}{@{\extracolsep{\fill}}lccccc}
        \toprule
          & Acc $\uparrow$ & Prec $\uparrow$ & Rec $\uparrow$ & F1 $\uparrow$ \\ 
        \midrule
        (a) C-VAE~\citep{harvey2022conditionalimagegenerationconditioning} & 0.84 $\pm$ 0.04 & 0.78 $\pm$ 0.04 & 0.69 $\pm$ 0.03 & 0.72 $\pm$ 0.03  \\
        (b) GNN~\citep{cui2023mgnngraphneuralnetworks} & 0.86 $\pm$ 0.06 & 0.86 $\pm$ 0.06 & 0.69 $\pm$ 0.05 & 0.73 $\pm$ 0.06 \\
        (c) Mesh~\citep{lionar2025treemeshgpt} & 0.84 $\pm$ 0.06 & 0.82 $\pm$ 0.11 &  0.69 $\pm$ 0.05 & 0.71 $\pm$ 0.05 \\
        (d) VisDiff~\citep{mahesh2026visdiff} & 0.86 $\pm$ 0.04 & 0.80 $\pm$ 0.04 & 0.74 $\pm$ 0.03 & 0.76 $\pm$ 0.03  \\
        (e) \ourmethodnewsmall~+~Optimization~(Ours) & \textbf{0.94 $\pm$ 0.03} & \textbf{0.91 $\pm$ 0.05} & \textbf{0.91 $\pm$ 0.05} & \textbf{0.91 $\pm$ 0.05} \\
        \bottomrule
    \end{tabularx}
    \vspace{0.3em}
    \caption{Baseline comparison: \textbf{Acc}: Accuracy, \textbf{Prec}: Precision, \textbf{Rec}: Recall, \textbf{F1}: F1-Score.}
    \label{tab:testset_res}
\end{table}

\subsection{Visibility Characterization}

We evaluate \ourmethodnewsmall{} on \ourproball{} by measuring coverage under local source-space perturbations. We sample a root noise vector \(S_0\) with variance \(\sigma^2\) for VisDiff and root triangle parameters \(T_0\) from the valid parameter ranges for \ourmethodnewsmall{}. Child samples are generated by perturbing their parent samples with Gaussian noise, with \(\sigma_p\) set to \(10\%\) of the corresponding sampling scale. We use \(T_d\in\{0.01,0.05\}\), \(T\in\{0.7,0.8\}\), \(b=2\), and \(d=5\), with \(T_d\) chosen from the minimum and maximum diversity observed among realizations of the same visibility graph in our dataset. Table~\ref{tab:coverage_metric_test_set} shows that \ourmethodnewsmall{} achieves substantially higher coverage than VisDiff, particularly at larger \(T_d\) and \(T\). We next analyze the source of this improvement. Table~\ref{tab:coverage_metric_approach_ablation} shows that inference-time optimization provides additional gains, while \ourmethodnewsmall{} already exceeds VisDiff without optimization. We then compare \ourmethodnewsmall{} with continuous SDF latent diffusion in Table~\ref{tab:coverage_metric_cont_visflow} to isolate the effect of polygon propagation. Perturbations of the latent-diffusion source produce few distinct polygons satisfying the target visibility constraint, whereas perturbations of the triangle source in \ourmethodnewsmall{} produce more polygons satisfying both acceptance criteria. These results indicate that polygon propagation is the primary source of the coverage improvement on \ourproball{}, while the \ourprob{} results show that the continuous SDF representation primarily improves visibility reconstruction.
\begin{table}[h!]
    \centering
    \small
    \begin{tabular}{cccc}
        \toprule
        F1 Threshold $T$ & Distance Threshold $T_d$ & VisDiff~\citep{mahesh2026visdiff}  & \ourmethodnewsmall{}~+~Optimization\\
        \midrule
        0.70 & 0.01 & 0.89 $\pm$ 0.28 & \textbf{1.00 $\pm$ 0.00}  \\
        0.70 & 0.05 &  0.68 $\pm$ 0.29 & \textbf{1.00 $\pm$ 0.01}   \\
        0.80 & 0.01 & 0.03 $\pm$ 0.16 & \textbf{0.95 $\pm$ 0.17} \\
        0.80 & 0.05 & 0.00 $\pm$ 0.00 & \textbf{0.95 $\pm$ 0.19}   \\
        \bottomrule
    \end{tabular}
    \vspace{0.3em}
    \caption{Coverage performance ($\uparrow$) under varying F1 threshold $T$ and distance threshold $T_d$: each value represents the fraction of nodes expanded by the method relative to the total number of nodes in the BFS search tree.}
    \label{tab:coverage_metric_test_set}
\end{table}

\vspace{-2.0em}
\subsection{Visibility Recognition}

\begin{figure}[h!]
    \centering
    \includegraphics[width=0.5\linewidth, height = 4cm]{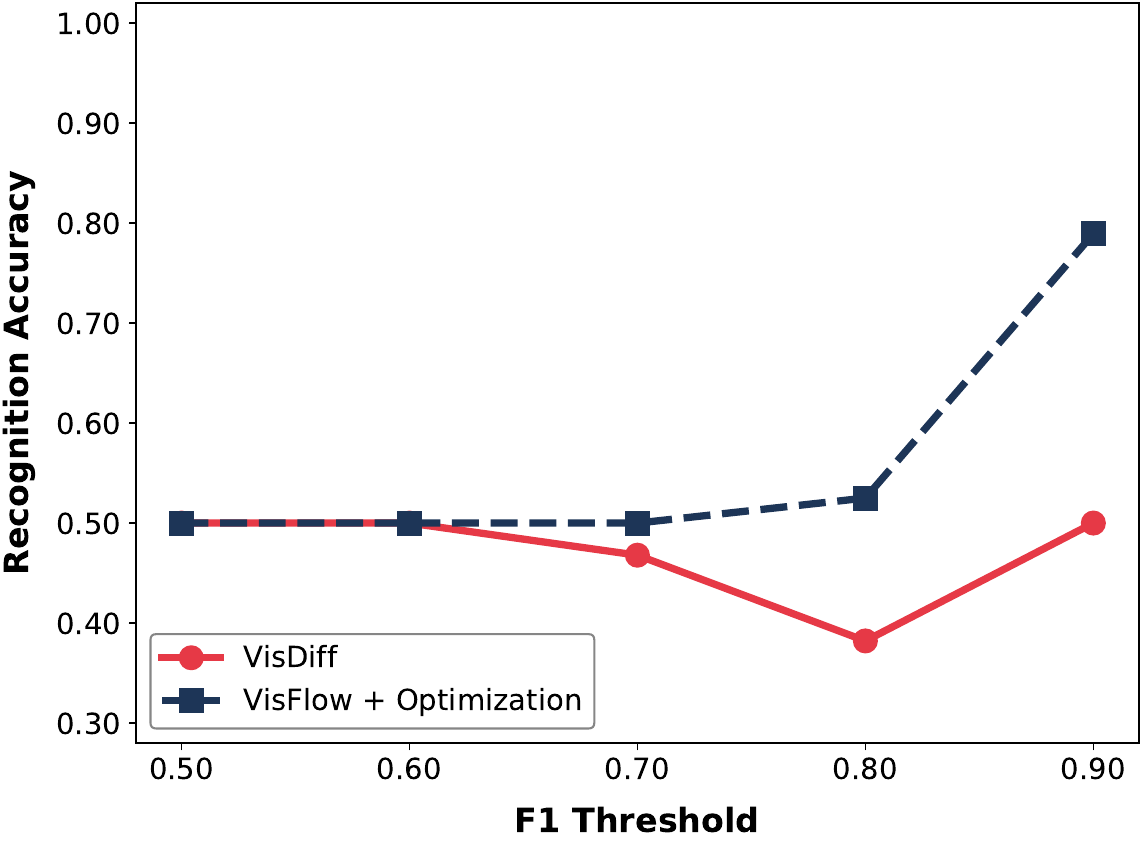}
    \caption{Recognition Performance: Accuracy vs F1 Threshold }
    \label{fig:visibility_recognition}
\end{figure}

We evaluate \ourmethodnewsmall{} on \ourprobrec{}. Following VisDiff, we construct the test set by adding holes to half of the polygons in the test split, producing visibility graphs that serve as proxies for invalid inputs. Polygons are generated using the same procedure as in \ourprob{}, and an input graph is classified as valid if the visibility F1-score of the selected polygon exceeds threshold TT. Figure~\ref{fig:visibility_recognition} compares \ourmethodnewsmall{} with VisDiff. VisDiff classifies all inputs as invalid, whereas \ourmethodnewsmall{} achieves 80\% recognition accuracy at a threshold near its mean F1-score on \ourprob{}. These results show that the improved reconstruction performance on \ourprob{} also translates to \ourprobrec{}.

\subsection{Contact-Graph Reconstruction}

\begin{figure*}[h!]
    \centering

    \begin{subfigure}{0.24\textwidth}
        \centering
        \includegraphics[width=\linewidth, trim=0cm 0cm 26cm 1cm, clip]{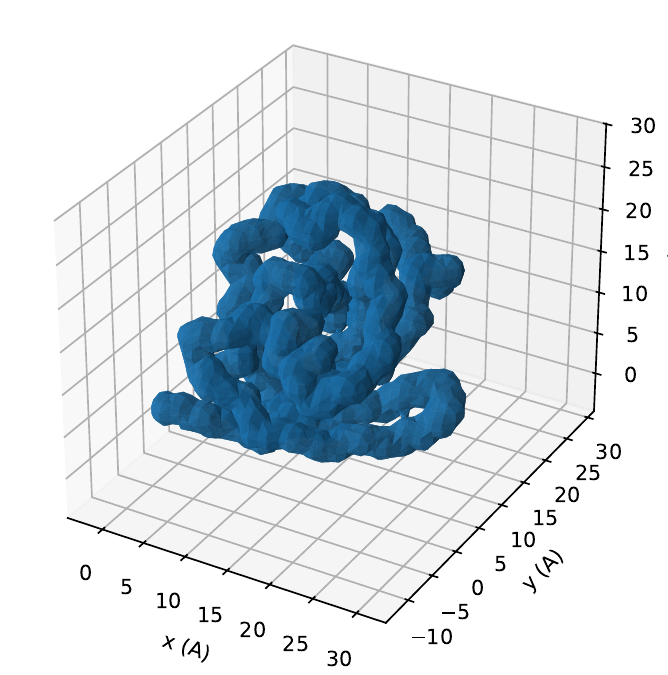}
        \caption{GT Protein Sample 1}
        \label{fig:subfig1}
    \end{subfigure}
    \hfill
    \begin{subfigure}{0.24\textwidth}
        \centering
        \includegraphics[width=\linewidth, trim=25cm 0cm 0cm 1cm, clip]{Figures/1a70A_gt_rfraw_rfpulchra}
        \caption{0.91}
        \label{fig:subfig2}
    \end{subfigure}
    \hfill
    \begin{subfigure}{0.24\textwidth}
        \centering
        \includegraphics[width=\linewidth, trim=0cm 0cm 26cm 1cm, clip]{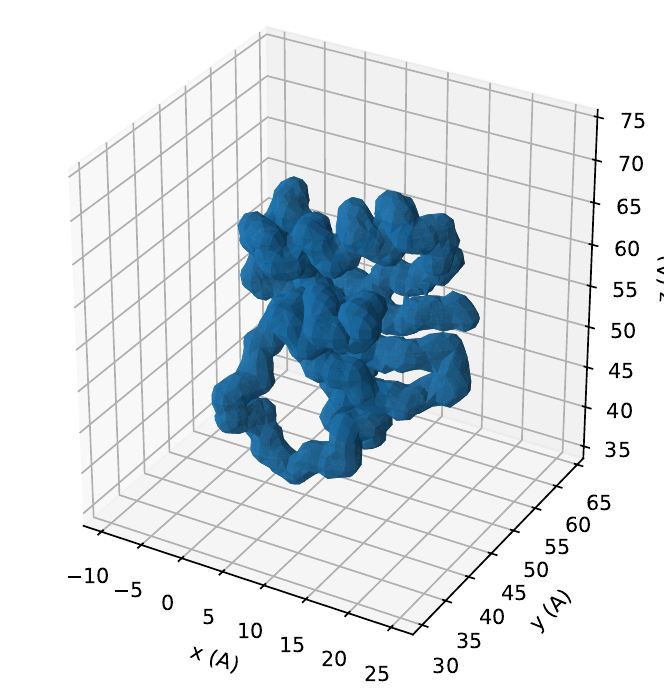}
        \caption{GT Protein Sample 2}
        \label{fig:subfig3}
    \end{subfigure}
    \hfill
    \begin{subfigure}{0.24\textwidth}
        \centering
        \includegraphics[width=\linewidth, trim=25cm 0cm 0cm 1cm, clip]{Figures/1abaA_gt_rfraw_rfpulchra.pdf}
        \caption{0.92}
        \label{fig:subfig4}
    \end{subfigure}

    \caption{\textbf{Contact-Graph Reconstruction.} Ground-truth protein structures are shown in (a) and (c), with their corresponding \ourmethodnewsmall{} reconstructions shown in (b) and (d), respectively. The labels report the contact-graph similarity between each ground-truth structure and its reconstruction.}
    \label{fig:protein}
\end{figure*}

We next evaluate whether the trajectory-supervision formulation of \ourmethodnewsmall{} extends beyond polygon reconstruction to protein contact-graph reconstruction. A protein contact graph is a binary adjacency matrix indicating whether residue pairs are spatially close. Since the polygon-formation trajectory does not represent protein conformational dynamics, we use targeted molecular dynamics~\citep{schlitter1994targeted} trajectories and reverse them to obtain random-coil-to-native reconstruction paths. We replace the polygon-specific encoder and decoder with a contact-graph encoder and a 3D coordinate decoder. The model is trained on ProteinNet~\citep{alquraishi2019proteinnet} using Kabsch-aligned mean-squared error~\citep{kabsch1976solution} together with a contact-graph reconstruction loss. We evaluate on FragFold150~\citep{kosciolek2014novo} by comparing the contact graph induced by the reconstructed structure with the target graph using accuracy, precision, recall, and F1-score. Table~\ref{tab:contact_graph_res} shows that \ourmethodnewsmall{} substantially outperforms the baselines. We omit VisDiff and MeshGPT because their architectures are specialized for 2D polygon and mesh representations and are not directly applicable to 3D protein contact-graph reconstruction. Figure~\ref{fig:protein} shows qualitative samples generated by \ourmethodnewsmall{}, and Appendix~\ref{app:contact_graph} provides architecture details.
\begin{table}[h!]
    \centering
    \begin{tabularx}{\textwidth}{@{\extracolsep{\fill}}lccccc}
        \toprule
          & Acc $\uparrow$ & Prec $\uparrow$ & Rec $\uparrow$ & F1 $\uparrow$ \\ 
        \midrule
        (a) C-VAE~\citep{harvey2022conditionalimagegenerationconditioning} & 0.79 $\pm$ 0.05 & 0.05 $\pm$ 0.02 & 0.32 $\pm$ 0.08 &  0.08 $\pm$ 0.03  \\
        (b) GNN~\citep{cui2023mgnngraphneuralnetworks} & 0.49 $\pm$ 0.13 & 0.06 $\pm$ 0.03 & 0.96 $\pm$ 0.03 & 0.11 $\pm$ 0.05 \\
        (c) \ourmethodnewsmall(Ours) & \textbf{0.99 $\pm$ 0.00} & \textbf{0.83 $\pm$ 0.08} & \textbf{0.89 $\pm$ 0.05} & \textbf{0.86 $\pm$ 0.06} \\
        \bottomrule
    \end{tabularx}
    \vspace{0.3em}
    \caption{Contact-Graph Reconstruction Performance Fragfold 150 Dataset: \textbf{Acc}: Accuracy, \textbf{Prec}: Precision, \textbf{Rec}: Recall, \textbf{F1}: F1-Score.}
    \label{tab:contact_graph_res}
\end{table}
\vspace{-2.0em}
\vspace{-2.0em}
\section{Conclusion}
We presented \ourmethodnew{}, an interpretable flow-matching model that combines continuous SDF representations with explicit polygon evolution for learning geometric realizations from visibility graphs. We show that the continuous SDF representation drives improvements in reconstruction accuracy, while trajectory-supervised polygon evolution leads to substantial gains in coverage and diversity of valid realizations. We introduce a new visibility-graph reconstruction dataset with \textbf{22\%} greater shape diversity than the state-of-the-art dataset, on which \ourmethodnew{} improves F1-score by \textbf{20\%} on \ourprob{} and coverage by \textbf{30\%} on \ourproball{}. We further show that the same trajectory-supervision formulation can be applied to protein structure reconstruction from contact graphs. However, our patch-based SDF representation currently assumes that shape information is distributed uniformly across fixed spatial patches. Future work will relax this assumption by introducing the patch radius as an additional parameter, following the original PatchNets~\citep{tretschk2020patchnets} formulation. We also plan to investigate contact-graph realizations as additional structural information for de novo protein generation with AlphaFold~\citep{jumper2021highly}.

\bibliography{iclr2027_conference}
\bibliographystyle{iclr2027_conference}

\newpage
\section{Appendix}
We first provide preleminary for our approach in \ref{app:prelim}. Next, we present the architectural details of the SDF Latent Generator in Section \ref{app:sdf_architecture} and the Flow Network in Section \ref{app:flow_architecture}. We then present the architecture and evaluation of the Visibility Prediction Network in Section \ref{app:vis_pred}. Next, we present additional Visibility Reconstruction results on the original VisDiff dataset in Section \ref{app:vis_rec}, followed by architectural and training details for contact-graph reconstruction in Section \ref{app:contact_graph}. We present ablation studies of the training losses and model components in Section \ref{app:ablation_study}. In Section \ref{app:vis_laten}, we compare VisFlow with continuous SDF latent diffusion to isolate the contribution of polygon propagation. We then compare the vertex prediction architectures of VisFlow and VisDiff in Section \ref{app:vertex_pred}. Finally, we analyze the statistics and diversity of our generated dataset in Section \ref{subsec:app_dataset_generation}.
\subsection{Preliminaries}
\label{app:prelim}
\par
\begin{figure}[h!]
    \centering
    \includegraphics[width=\textwidth{}, height = 3.5cm]{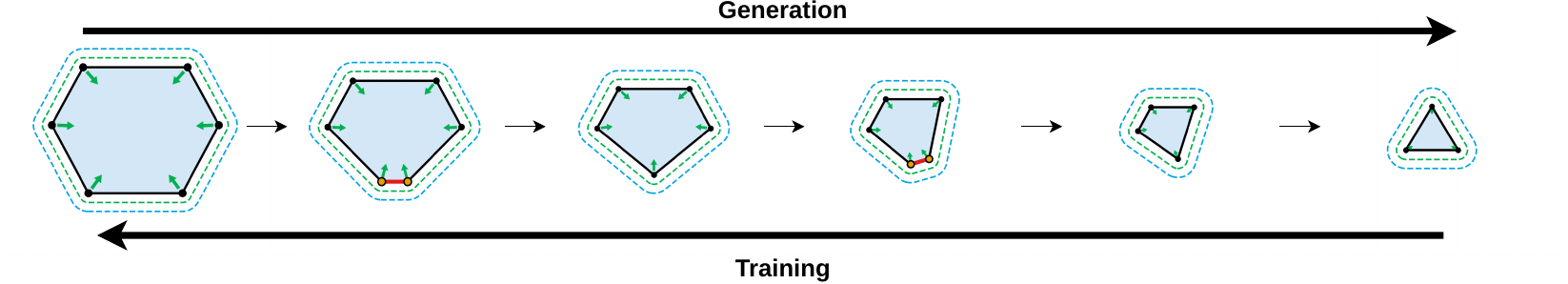}
    \caption{
    \textbf{Grassfire trajectory for polygon formation.} Ground-truth trajectories are constructed by applying grassfire flow to collapse the polygon along local SDF-normal directions until only a triangle remains. The arrows indicate the direction of vertex motion. The yellow vertices mark an edge-collapse event, in which two neighboring vertices meet as the edge between them shrinks to zero length, after which the polygon continues with one fewer vertex. For training, we reverse the trajectory so that polygon formation proceeds from the triangle to the complete polygon. The intermediate grassfire states supervise the corresponding points along the latent trajectory.
    }
    \label{fig:grassfire}
\end{figure}

\textbf{Grassfire Evolution:}
Grassfire defines a deterministic inward evolution of the polygon boundary in which vertices move along local SDF-normal directions. An edge-collapse event occurs as two neighboring vertices meet and the intervening edge shrinks to zero length, reducing the number of polygon vertices. Figure~\ref{fig:grassfire} illustrates the evolution and edge-collapse process.

\par
\begin{figure*}[h]
    \centering
    \begin{subfigure}[b]{0.30\textwidth}
        \centering
        \includegraphics[width=\linewidth, height = 3cm]{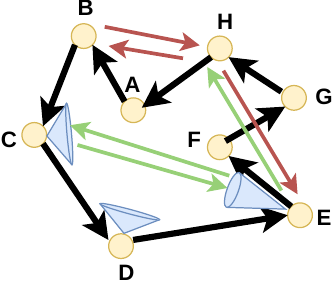}
        \caption{In-cone condition}
        \label{fig:incone}
    \end{subfigure}
    \hfill
    \begin{subfigure}[b]{0.45\textwidth}
        \centering
        \includegraphics[width=\linewidth]{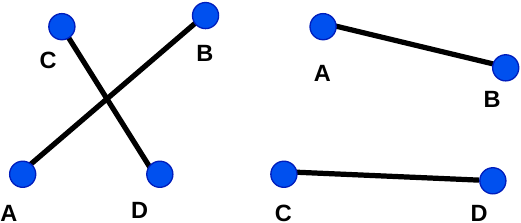}
        \caption{Intersection condition}
        \label{fig:intersection}
    \end{subfigure}
\caption{\textbf{Visibility predicates.}
(a) Two polygon vertices can be visible only if their connecting segment lies within the in-cone region at both endpoints. At each endpoint, the two incident polygon edges define an angular region, and the portion of this region lying inside the polygon forms the local in-cone region, shown in blue. Green arrows indicate that the segment from the other vertex enters through the in-cone region, while red arrows indicate that it enters from outside. Thus, $\overline{BH}$ fails at both endpoints, $\overline{HE}$ satisfies the condition only at $H$, and $\overline{CE}$ satisfies it at both endpoints. (b) The intersection predicate determines whether a candidate segment intersects a non-incident polygon edge.}
    \label{fig:visibility_predicates}
\end{figure*}

\textbf{Visibility Formulation:}
Two polygon vertices are visible if the segment connecting them lies entirely within the polygon. Visibility decomposes into an in-cone condition at both endpoints and an intersection condition with non-incident polygon edges~\citep{o1998computational}. Visibility between vertices $v_i$ and $v_j$ is defined as
\[
V_{ij}
=
I_{\mathrm{InCone}}(v_i,v_j)
\left(1-I_{\mathrm{Intersect}}(v_i,v_j)\right),
\]
where $I_{\mathrm{InCone}}\in{0,1}$ indicates whether $\overline{v_i v_j}$ lies within the valid visibility cone at both endpoints, and $I_{\mathrm{Intersect}}\in{0,1}$ indicates whether it intersects any non-incident polygon edge. Figure~\ref{fig:visibility_predicates} illustrates both predicates. The Visibility Prediction Network follows the same decomposition.

\subsection{SDF Latent Generator Architecture}
\label{app:sdf_architecture}
In this section, we present the architectural details fo SDF Ltent Generator. We represent each polygon using a spatially structured SDF latent representation. The domain \([-1,1]^2\) is uniformly partitioned into a \(6\times6\) grid, yielding \(N=36\) spatially localized patches. Each SDF sample is first mapped to a feature representation using PointNet~\citep{qi2017pointnet} and assigned to the patch whose center is nearest to the sample location. Within each patch, the sample features are aggregated using a Gaussian-decayed inverse-distance-weighted average based on their distance to the patch center, producing one latent vector per patch. We use a fixed uniform partition rather than adaptive patch placement to maintain a consistent spatial decomposition across shapes and avoid the additional complexity of adaptive patch optimization~\citep{tretschk2020patchnets}. The polygon decoder uses \(Q=50\) learnable vertex queries to recover the polygon vertex locations from the patch latents. The visibility graph \(G\) is represented using its upper-triangular entries and processed by an MLP that produces one visibility embedding for each vertex. Each visibility embedding is concatenated with the corresponding vertex query, and the resulting query representations are processed by a Transformer decoder that cross-attends to the patch latents to predict the polygon vertex locations.

\subsection{Flow Network Architecture}
\label{app:flow_architecture}

In this section, we present the architectural details fo Flow Network. The flow network operates on a spatially structured SDF latent representation consisting of \(36\) patch latents associated with a uniform \(6\times6\) partition of the domain \([-1,1]^2\). The patch latents are arranged according to their spatial locations to form a \(6\times6\) latent feature map. We parameterize the velocity field using a time-conditioned U-Net following the architecture of Latent Diffusion~\citep{rombach2022high}.

The target visibility graph \(G\) is represented by its upper-triangular entries. A visibility encoder processes this global graph representation using an MLP and produces one visibility embedding for each vertex. The resulting per-vertex visibility embeddings condition the U-Net through cross-attention, allowing the predicted velocity field to depend on both the current latent state and the target visibility graph.
\subsection{Visibility Prediction Network}
\label{app:vis_pred}
\begin{figure}[h!]
    \centering
    \includegraphics[width=\columnwidth]{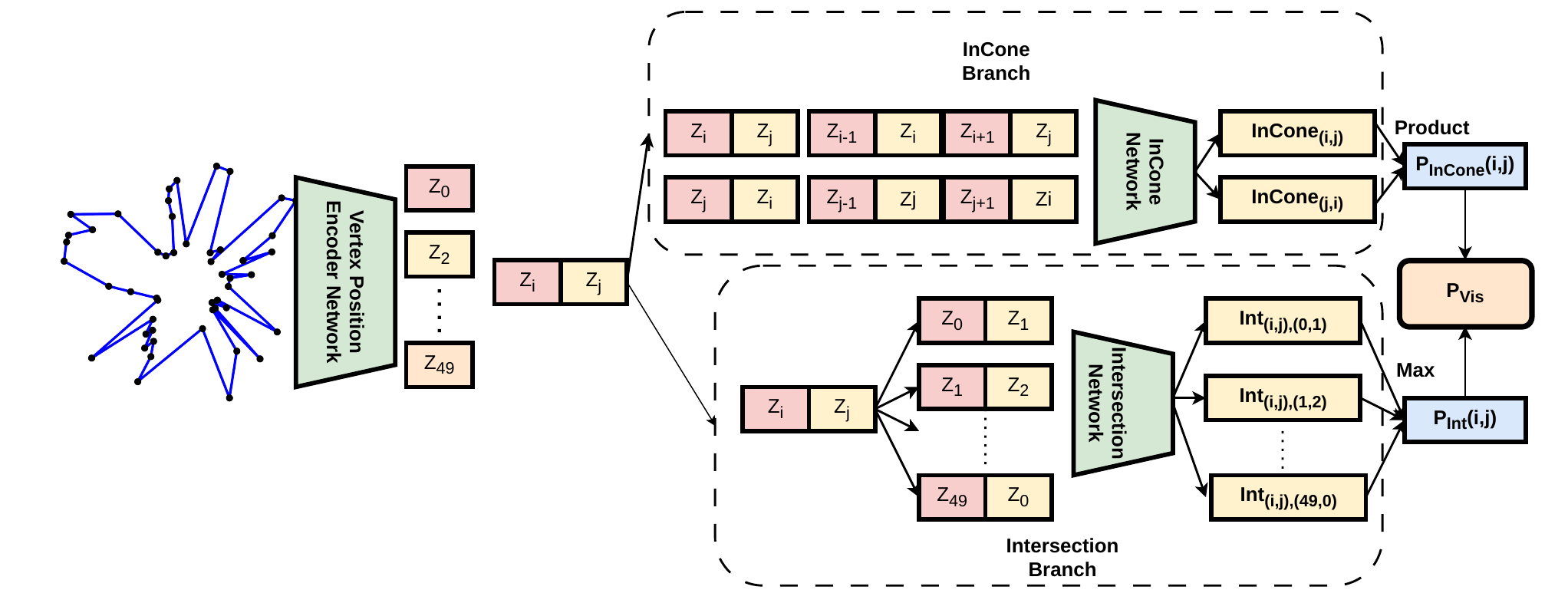}
    \caption{\textbf{Architecture of the Visibility Prediction Network.} Polygon vertices are first mapped to positional embeddings by a shared vertex encoder. For a candidate segment $(v_i,v_j)$, the \textbf{In-Cone Branch} combines the candidate-segment representation with the neighboring polygon edges at each endpoint and predicts an in-cone probability for both endpoints. The two endpoint probabilities are multiplied to obtain $P_{\mathrm{InCone}}$. The \textbf{Intersection Branch} compares the candidate segment with each non-incident polygon boundary edge and predicts pairwise intersection probabilities, which are aggregated using a max operation to obtain $P_{\mathrm{Intersect}}$. The branch outputs are combined as $P_{\mathrm{Vis}}=P_{\mathrm{InCone}}(1-P_{\mathrm{Intersect}})$ to obtain the final visibility probability for $(v_i,v_j)$.}
    \label{fig:verifer_architecture}
\end{figure}

In this section, we present the detailed architecture of the Visibility Prediction Network and evaluate its performance on ground-truth polygons from the test split of our dataset. Figure~\ref{fig:verifer_architecture} illustrates the network architecture. We assess its ability to recover visibility relationships through a comparison with a conventional transformer architecture in Table~\ref{tab:verifier_performance}. The results show that our network perfectly reconstructs the visibility graphs of the ground-truth polygons in the test set and substantially outperforms the transformer baseline.
\begin{table}[h!]
    \centering
    \begin{tabularx}{\textwidth}{@{\extracolsep{\fill}}lccccc}
        \toprule
          & Acc $\uparrow$ & Prec $\uparrow$ & Rec $\uparrow$ & F1 $\uparrow$ \\ 
        \midrule
        (a) Visibility Verifier Network & 1.00 $\pm$ 0.00 & 1.00 $\pm$ 0.00 & 1.00 $\pm$ 0.00 & 1.00 $\pm$ 0.00  \\
        (b) Transformer Architecture & 0.91 $\pm$ 0.02 & 0.89 $\pm$ 0.03 & 0.76 $\pm$ 0.05 & 0.82 $\pm$ 0.04 \\
    \bottomrule
    \end{tabularx}
    \vspace{0.3em}
    \caption{Vertex Prediction comparison: \textbf{Acc}: Accuracy, \textbf{Prec}: Precision, \textbf{Rec}: Recall, \textbf{F1}: F1-Score.}
    \label{tab:verifier_performance}
\end{table}

\subsection{\ourprob{}}
\label{app:vis_rec}
In this section, we additionally evaluate \ourmethodnewsmall{} on the original VisDiff dataset~\citep{mahesh2026visdiff} and compare its performance with existing baselines. Table \ref{tab:testset_visdiff_res} shows the comparison between \ourmethodnewsmall{} and baselines. \ourmethodnewsmall{} outperforms all baselines showing that the improvements of \ourmethodnewsmall{} are also retained on the original VisDiff dataset.

\begin{table}[h!]
    \centering
    \begin{tabularx}{\textwidth}{@{\extracolsep{\fill}}lccccc}
        \toprule
          & Acc $\uparrow$ & Prec $\uparrow$ & Rec $\uparrow$ & F1 $\uparrow$ \\ 
        \midrule
        (a) C-VAE~\citep{harvey2022conditionalimagegenerationconditioning} & 0.74 & 0.72 & 0.70 & 0.70  \\
        (b) GNN~\citep{cui2023mgnngraphneuralnetworks} & 0.73 & 0.79 & 0.69 & 0.67 \\
        (c) Mesh~\citep{lionar2025treemeshgpt} & 0.75 & 0.74 &  0.72 & 0.71 \\
        (d) VisDiff~\citep{mahesh2026visdiff} & 0.92 & 0.91 & 0.91 & 0.91  \\
        (e) \ourmethodnewsmall~+~Optimization~(Ours) & \textbf{0.97} & \textbf{0.98} & \textbf{0.96} & \textbf{0.97} \\
        \bottomrule
    \end{tabularx}
    \vspace{0.3em}
    \caption{Baseline comparison VisDiff Dataset: \textbf{Acc}: Accuracy, \textbf{Prec}: Precision, \textbf{Rec}: Recall, \textbf{F1}: F1-Score.}
    \label{tab:testset_visdiff_res}
\end{table}

\subsection{Contact-Graph Reconstruction}
\label{app:contact_graph}

In this section, we present the architecture details for the contact-graph reconstruction problem. We use the same trajectory-supervised latent-flow formulation as \ourmethodnewsmall{}, while replacing the polygon-specific components with a protein SDF encoder, a backbone decoder, and a contact-graph encoder. We define the protein contact graph \(G\) as a binary adjacency matrix over residues, where \(G_{ij}=1\) if the C$\alpha$ atoms of residues \(i\) and \(j\) are within \(8\,\text{\AA}\), and \(G_{ij}=0\) otherwise.

Protein backbones are represented using continuous SDF samples computed from the N, C$\alpha$, and C atoms. The SDF encoder uses \(N=64\) patch latent tokens of dimension 128, forming a \(128\times8\times8\) latent representation. The backbone decoder predicts N/C$\alpha$/C coordinates for proteins containing up to 270 residues and is conditioned on both the amino-acid sequence and the target contact graph. We use a maximum length of 270 residues because it covers the proteins in the training and evaluation datasets while maintaining a fixed-size representation. The contact graph is represented as a padded \(270\times270\) matrix and encoded row-wise using an MLP followed by a six-layer Transformer. The sequence embeddings and graph features are jointly processed by a six-layer Transformer decoder with hidden dimension 128 and eight attention heads.

The SDF latent representation is trained using the same loss formulation as in polygon generation, with a reconstruction loss \(\mathcal{L}_{\mathrm{rec}}\), alignment loss \(\mathcal{L}_{\mathrm{ali}}\), and latent regularization loss \(\mathcal{L}_{\mathrm{reg}}\). For protein reconstruction, \(\mathcal{L}_{\mathrm{rec}}\) consists of a Kabsch-aligned backbone coordinate reconstruction loss \(\mathcal{L}_{\mathrm{coord}}\)~\citep{kabsch1976solution} and a contact-graph consistency loss \(\mathcal{L}_{\mathrm{contact}}\). Given the predicted C$\alpha$ coordinates \(\hat{\mathbf{x}}_i\), we compute the pairwise distances as
\begin{equation}
    d_{ij}=\|\hat{\mathbf{x}}_i-\hat{\mathbf{x}}_j\|_2,
\end{equation}
and define the corresponding contact logits as
\begin{equation}
    \ell_{ij}=\frac{8-d_{ij}}{\tau},
    \qquad \tau=0.5.
\end{equation}
We define \(\mathcal{L}_{\mathrm{contact}}\) as a weighted binary cross-entropy loss between the predicted contacts and the target contact graph for residue pairs satisfying \(|i-j|\geq6\), which excludes short-range sequence contacts and focuses the loss on long-range structural relationships. We use weights \(0.05\) for contacts and \(0.95\) for non-contacts. The reconstruction loss is
\begin{equation}
    \mathcal{L}_{\mathrm{rec}}
    =
    \mathcal{L}_{\mathrm{coord}}
    +
    \lambda_{\mathrm{contact}}\mathcal{L}_{\mathrm{contact}},
\end{equation}
and the overall latent objective is
\begin{equation}
    \mathcal{L}_{\mathrm{latent}}
    =
    \mathcal{L}_{\mathrm{rec}}
    +
    \lambda_{\mathrm{ali}}\mathcal{L}_{\mathrm{ali}}
    +
    \lambda_{\mathrm{reg}}\mathcal{L}_{\mathrm{reg}}.
\end{equation}
We set \(\lambda_{\mathrm{contact}}=0.01\), \(\lambda_{\mathrm{ali}}=1.0\), and \(\lambda_{\mathrm{reg}}=10^{-3}\).

The SDF encoder and backbone decoder are frozen once the latent representation has been trained. We then train the contact-graph-conditioned Rectified Flow using the same formulation as in polygon generation, with the target contact graph providing the conditioning signal to the latent U-Net. The Rectified Flow model is trained using AdamW with a learning rate of \(10^{-4}\) and weight decay of \(10^{-4}\), and inference is performed using 60 Euler steps.
\subsection{Ablation Study}
\label{app:ablation_study}
In this section, we present an ablation study of the training loss on the \ourprob{} problem. We also examine the contribution of each module in \ourmethodnewsmall{} to performance on \ourprob{} and \ourproball{}. Table~\ref{tab:loss_ablation} reports the effect of adding the visibility loss on \ourprob{}. The results show that incorporating the visibility loss substantially improves reconstruction performance. Tables~\ref{tab:approach_ablation_study} and~\ref{tab:coverage_metric_approach_ablation} report the effects of adding each module on \ourprob{} and \ourproball{}, respectively. The results show that incorporating optimization into \ourmethodnewsmall{} improves performance on both problems.
\begin{table}[h!]
    \centering
    \begin{tabularx}{\textwidth}{@{\extracolsep{\fill}}lccccc}
        \toprule
          & Acc $\uparrow$ & Prec $\uparrow$ & Rec $\uparrow$ & F1 $\uparrow$ \\ 
        \midrule
        (a) MSE Loss & 0.87 $\pm$ 0.05 & 0.81 $\pm$ 0.04 & 0.75 $\pm$ 0.03 & 0.77 $\pm$ 0.03   \\
        (b) MSE + Visibility Loss & \textbf{0.89 $\pm$ 0.02} & \textbf{0.84 $\pm$ 0.04} & \textbf{0.82 $\pm$ 0.04} & \textbf{0.83 $\pm$ 0.04}  \\
        \bottomrule
    \end{tabularx}
    \vspace{0.3em}
    \caption{Loss Ablation Study: \textbf{Acc}: Accuracy, \textbf{Prec}: Precision, \textbf{Rec}: Recall, \textbf{F1}: F1-Score.}
    \label{tab:loss_ablation}
\end{table}

\begin{table}[h!]
    \centering
    \begin{tabularx}{\textwidth}{@{\extracolsep{\fill}}lccccc}
        \toprule
          & Acc $\uparrow$ & Prec $\uparrow$ & Rec $\uparrow$ & F1 $\uparrow$ \\ 
        \midrule
        (a) \ourmethodnewsmall & 0.89 $\pm$ 0.02 & 0.84 $\pm$ 0.04 & 0.82 $\pm$ 0.04 & 0.83 $\pm$ 0.04 \\
        (b) \ourmethodnewsmall~+~Optimization & \textbf{0.94 $\pm$ 0.03} & \textbf{0.91 $\pm$ 0.05} & \textbf{0.91 $\pm$ 0.05} & \textbf{0.91 $\pm$ 0.05} \\
        (c) Optimization & 0.88 $\pm$ 0.04 & 0.81 $\pm$ 0.07 & 0.88 $\pm$ 0.04 & 0.83 $\pm$ 0.06 \\
        \bottomrule
    \end{tabularx}
    \vspace{0.3em}
    \caption{Approach Ablation Study Reconstruction Problem: \textbf{Acc}: Accuracy, \textbf{Prec}: Precision, \textbf{Rec}: Recall, \textbf{F1}: F1-Score.}
    \label{tab:approach_ablation_study}
\end{table}

\begin{table}[h!]
    \centering
    \small
    \begin{tabular}{cccc}
        \toprule
        F1 Threshold $T$ & Distance Threshold $T_d$ & \ourmethodnewsmall{}~+~Optimization & \ourmethodnewsmall{}\\
        \midrule
        0.70 & 0.01 & \textbf{1.00 $\pm$ 0.00} & 1.00 $\pm$ 0.00  \\
        0.70 & 0.05 &  \textbf{1.00 $\pm$ 0.01} & 0.76 $\pm$ 0.21   \\
        0.80 & 0.01 & \textbf{0.95 $\pm$ 0.17} & 0.51 $\pm$ 0.47 \\
        0.80 & 0.05 & \textbf{0.95 $\pm$ 0.19} & 0.36 $\pm$ 0.35   \\
        \bottomrule
    \end{tabular}
    \vspace{0.3em}
    \caption{Approach Ablation Study Characterization Problem: each value represents the fraction of nodes expanded by the method relative to the total number of nodes in the BFS search tree.}
    \label{tab:coverage_metric_approach_ablation}
\end{table}
\subsection{How does \ourmethodnewsmall{} compare with latent diffusion?}
\label{app:vis_laten}
In this section, we examine whether \ourmethodnewsmall{} offers advantages over a conventional latent diffusion model operating on continuous SDF latents. We compare \ourmethodnewsmall{} with \ourmethodsmall{}, which performs diffusion in the SDF latent space without polygon propagation. We evaluate both methods on \ourprob{} and \ourproball{}. Table~\ref{tab:cont_visdiff_vs_visflow} presents the results on \ourprob{}. The results show that polygon propagation provides no meaningful improvement in reconstruction performance. Table~\ref{tab:coverage_metric_cont_visflow} presents the corresponding results on \ourproball{}. Here, \ourmethodnewsmall{} achieves substantial gains over \ourmethodsmall{}. The findings highlight the contribution of polygon propagation to characterization by enabling more effective exploration of distinct polygons consistent with a target visibility graph.

\begin{table}[h!]
    \centering
    \begin{tabularx}{\textwidth}{@{\extracolsep{\fill}}lccccc}
        \toprule
          & Acc $\uparrow$ & Prec $\uparrow$ & Rec $\uparrow$ & F1 $\uparrow$ \\ 
        \midrule
                (a) \ourmethodsmall & 0.88 $\pm$ 0.02 & 0.82 $\pm$ 0.03 & 0.83 $\pm$ 0.03 & 0.82 $\pm$ 0.03  \\
        (b) \ourmethodnewsmall & 0.89 $\pm$ 0.02 & 0.84 $\pm$ 0.04 & 0.82 $\pm$ 0.04 & 0.83 $\pm$ 0.04 \\

        (c) \ourmethodsmall~+~Optimization & \textbf{0.94 $\pm$ 0.02} & \textbf{0.91 $\pm$ 0.03} & \textbf{0.91 $\pm$ 0.03} & \textbf{0.91 $\pm$ 0.03}  \\
        (d) \ourmethodnewsmall~+~Optimization & \textbf{0.94 $\pm$ 0.03} & \textbf{0.91 $\pm$ 0.05} & \textbf{0.91 $\pm$ 0.05} & \textbf{0.91 $\pm$ 0.05} \\
        \bottomrule
    \end{tabularx}
    \vspace{0.3em}
    \caption{Approach Ablation Study: \textbf{Acc}: Accuracy, \textbf{Prec}: Precision, \textbf{Rec}: Recall, \textbf{F1}: F1-Score.}
    \label{tab:cont_visdiff_vs_visflow}
\end{table}

\begin{table}[h!]
    \centering
    \small
    \begin{tabular}{cccc}
        \toprule
        F1 Threshold $T$ & Distance Threshold $T_d$ & \ourmethodsmall{}  & \ourmethodnewsmall{} 
        \\
        \midrule
        0.70 & 0.01 & 0.05 $\pm$ 0.10 & \textbf{1.00 $\pm$ 0.00} 
        \\
        0.70 & 0.05 & 0.01 $\pm$ 0.00 & \textbf{0.76 $\pm$ 0.21}
        \\
        0.80 & 0.01 & 0.05 $\pm$ 0.10 & \textbf{0.51 $\pm$ 0.47} 
        \\
        0.80 & 0.05 & 0.01 $\pm$ 0.00 & \textbf{0.36 $\pm$ 0.35} 
        \\
        \bottomrule
    \end{tabular}
    \vspace{0.3em}
    \caption{Coverage performance ($\uparrow$) under varying F1 threshold $T$ and distance threshold $T_d$: each value represents the fraction of nodes expanded by the method relative to the total number of nodes in the BFS search tree.}
    \label{tab:coverage_metric_cont_visflow}
\end{table}


\subsection{Vertex Prediction Architecture}
\label{app:vertex_pred}
In this section, we compare the vertex prediction architectures of \ourmethodnewsmall{} and VisDiff. We provide each architecture with the ground-truth SDF and the corresponding visibility graph to assess its ability to recover polygon vertices. Table~\ref{tab:GT_SDF_Vertex_Pred} presents the comparison. The results show that \ourmethodnewsmall{} recovers polygons more reliably than VisDiff even when both methods receive ground-truth SDFs. The comparison highlights the limitations of VisDiff's coarse SDF resolution for preserving the geometric detail required for accurate vertex recovery, while the continuous SDF representation in \ourmethodnewsmall{} supports more accurate polygon reconstruction.
\begin{table}[h!]
    \centering
    \begin{tabularx}{\textwidth}{@{\extracolsep{\fill}}lccccc}
        \toprule
          & Acc $\uparrow$ & Prec $\uparrow$ & Rec $\uparrow$ & F1 $\uparrow$ \\ 
        \midrule
        (a) VisDiff~\citep{mahesh2026visdiff} & 0.84 $\pm$ 0.05 & 0.78 $\pm$ 0.04 & 0.72 $\pm$ 0.03 & 0.74 $\pm$ 0.03  \\
        (b) \ourmethodsmall & 0.88 $\pm$ 0.02 & 0.81 $\pm$ 0.04 & 0.82 $\pm$ 0.03 & 0.82 $\pm$ 0.03  \\
        (c) \ourmethodnewsmall & \textbf{0.89 $\pm$ 0.02} & \textbf{0.82 $\pm$ 0.04} & \textbf{0.83 $\pm$ 0.04} & \textbf{0.83 $\pm$ 0.04}  \\
        \bottomrule
    \end{tabularx}
    \vspace{0.3em}
    \caption{Vertex Prediction comparison: \textbf{Acc}: Accuracy, \textbf{Prec}: Precision, \textbf{Rec}: Recall, \textbf{F1}: F1-Score.}
    \label{tab:GT_SDF_Vertex_Pred}
\end{table}
\subsection{Dataset Generation}
\label{subsec:app_dataset_generation}
\begin{figure}[h!]
    \centering

    \begin{subfigure}[b]{0.48\columnwidth}
        \centering
        \includegraphics[width=\linewidth]{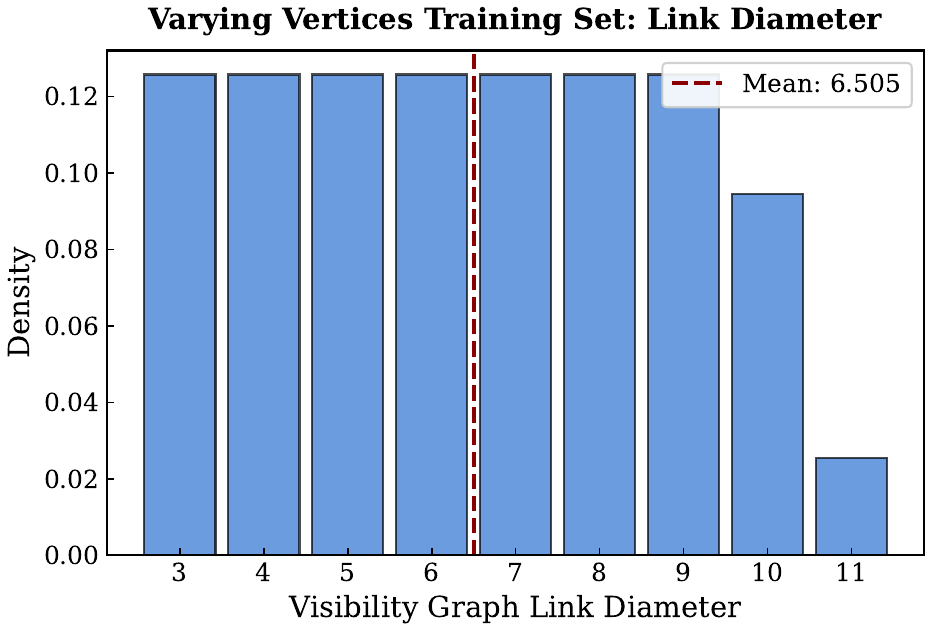}
        \caption{Link Diameter: Train Distribution}
        \label{fig:link_diameter}
    \end{subfigure}
    \hfill
    \begin{subfigure}[b]{0.48\columnwidth}
        \centering
        \includegraphics[width=\linewidth]{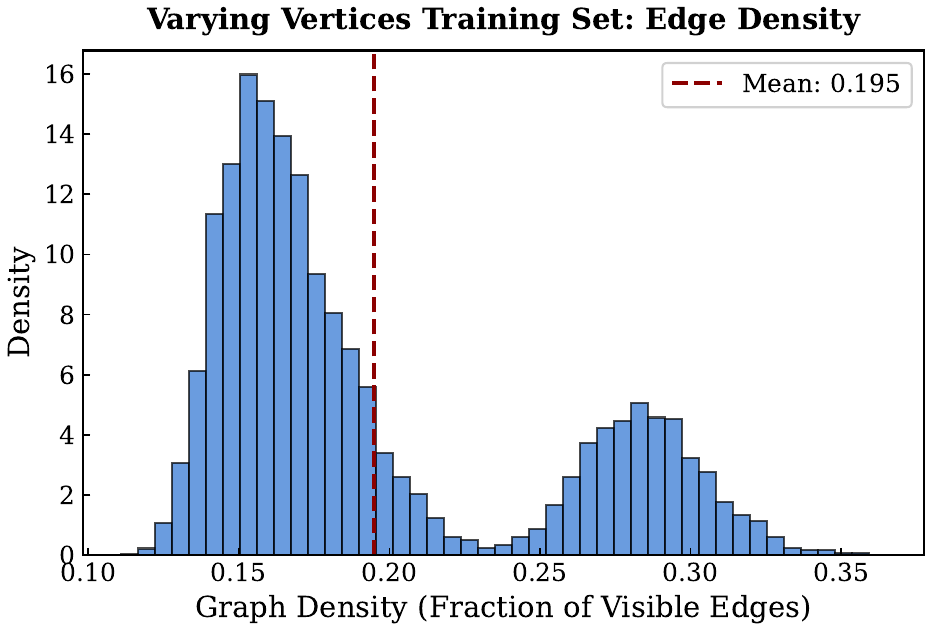}
        \caption{Graph Connectivity: Train Distribution}
        \label{fig:connectivity}
    \end{subfigure}

    \caption{Dataset Statistics: Distribution of (a) polygon link diameter and (b) visibility-graph edge density across polygons in the \ourmethodnewsmall{} after resampling based on link diameter.}
    \label{fig:dataset_statistics}
\end{figure}

Earlier datasets contain limited variation in polygon concavity. VisDiff broadens the range of concave geometries, but its coarse grid-based SDF limits the geometries that can be represented accurately. We therefore construct a dataset with greater geometric and visibility-graph diversity. We generate a large pool of random polygons using the 2-opt move algorithm~\citep{auer1996rpg}, normalize them to \([-1,1]^2\), order their vertices counterclockwise, and canonicalize the ordering by choosing the bottom-left vertex as the first vertex. We compute their visibility graphs and select 10,000 polygons with unique visibility graphs, sampling across link-diameter values to cover a broad range of concavities. For each polygon, we generate 20 additional instances through vertex perturbations that preserve its visibility graph, providing geometrically distinct realizations of the same visibility structure for the one-to-many \ourproball{} problem. The test set uses visibility graphs excluded from training and sampled from the remaining polygon pool, ensuring evaluation on unseen visibility structures. We analyze the generated dataset using link diameter and visibility-graph edge density to characterize its geometric and structural variation. Figure~\ref{fig:dataset_statistics} shows the training distributions of these statistics after resampling across link-diameter values. We further compare the range of polygon concavities in our dataset with existing datasets using link diameter. As shown in Table~\ref{tab:other_datasets}, our dataset has a larger average link diameter and covers a wider range of link-diameter values than the compared datasets. In particular, the maximum link diameter increases from $9.0$ in VisDiff to $11.0$ in our dataset, while the higher average link diameter indicates greater representation of highly concave polygons. These statistics reflect the broader range of polygon structures included in our dataset.

\begin{table}[h!]
    \centering
    \begin{tabularx}{\textwidth}{@{\extracolsep{\fill}}lccc}
        \toprule
        Baselines & Link Diameter (Avg / Std) & Min & Max \\ 
        \midrule
        MNIST~\citep{lecun1998mnist} & 1.83 / 0.50 & 1.0 & 4.1 \\
        COCO 2017~\citep{lin2014microsoft} & 1.32 / 0.25 & 1.0 & 3.823 \\
        VisDiff~\citep{mahesh2026visdiff} & 4.4 / 2.2 & 1.0 & 9.0 \\
        \ourmethodnewsmall{} & 6.5 / 2.32 & 1.0 & 11.0 \\
        \bottomrule
    \end{tabularx}
    \vspace{0.3em}
    \caption{Real-world dataset comparison showing the diversity (in link diameter) between our dataset and standard benchmarks.}
    \label{tab:other_datasets}
\end{table}
\label{subsec:testdataset_gen}

\end{document}